\documentclass{article}
\usepackage{spconf,amsmath,graphicx}
\usepackage{amssymb}

\usepackage{amsmath}
\usepackage{algorithm}
\usepackage[noend]{algpseudocode}
\usepackage{multirow}
\usepackage{url}
\newcommand{\norm}[1]{\left\lVert#1\right\rVert}
\title{Speaker verification using end-to-end adversarial language adaptation}
\name{Johan Rohdin$^{1*}$, Themos Stafylakis$^{2*}$, Anna Silnova$^1$, Hossein Zeinali$^1$, Luk\'a\v{s} Burget$^1$, Old\v{r}ich Plchot$^1$}
\address{$^1$Brno University of Technology, Speech@FIT, Brno, Czech Republic \\
\texttt{\{rohdin,isilnova,zeinali,burget,plchot\}@fit.vutbr.cz} \\
$^2$Omilia - Conversational Intelligence, Athens, Greece\\
\texttt{tstafylakis@omilia.com} \\
\thanks{*Equal Contribution. \newline
 This project has received funding from the European union's Horizon 2020 research and innovation programme under the Marie Sklodowska-Curie and it is co-financed by the South Moravian Region under grant agreement No. 665860. The project was also supported by the Czech Science Foundation under project No. GJ17-23870Y.
}
}

\begin{document}
%\ninept
%
\maketitle
\begin{abstract}
In this paper we investigate the use of adversarial domain adaptation for addressing the problem of language mismatch between speaker recognition corpora. In the context of speaker verification, adversarial domain adaptation methods aim at minimizing certain divergences between the distribution that the utterance-level features follow (i.e. speaker embeddings) when drawn from source and target domains (i.e. languages), while preserving their capacity in recognizing speakers. Neural architectures for extracting utterance-level representations enable us to apply adversarial adaptation methods in an end-to-end fashion and train the network jointly with the standard cross-entropy loss. We examine several configurations, such as the use of (pseudo-)labels on the target domain as well as domain labels in the feature extractor, and we demonstrate the effectiveness of our method on the challenging NIST SRE16 and SRE18 benchmarks. 
\end{abstract}
\section{Introduction}
\label{sec:intro}
The need for domain adaptation (DA) arises in cases when the target domain data is insufficient (and possibly unlabeled) for training a model from scratch and therefore source domain data (assumed labeled and sufficient for training a model) should be leveraged as well. The core idea behind DA is that the knowledge distilled from the source domain can be transferred to the target domain, despite the differences in the marginal distributions of the two domains. Conventional approaches to DA, such as fine-tuning a source domain model to the target domain data may fail in many settings due to the target data being weakly-labeled or even unlabeled. 

DA methods for speaker verification are of particular interest, as for many real-world applications large amounts of target domain labeled data are rarely available. Hence, for training state-of-the-art models which require several thousand of training utterances, one should resort to large out-of-domain corpora and use the small and possibly unlabeled target domain datasets for language, channel or other types of adaptation. In order to promote further research in DA methods, MIT-LL and NIST has organized 3 evaluations (namely the MIT-LL DA challenge, DAC-2013, NIST SRE16, and the recent NIST SRE18) with the two latter focused primarily on language adaptation. Several DA methods were introduced as part of those evaluations, the majority of which approach the problem as a transformation of fixed utterance-level representations, such as i-vectors.

In this article we examine the use of the recently emerged adversarial DA methods. Adversarial DA methods employ Generative Adversarial Networks (GANs) as a means to reduce the mismatch between source and target domains \cite{Ganin_JMLR2016}. Different from \cite{wang2018unsupervised}, where an adversarial architecture is proposed for i-vector adaptation and tested for mildly mismatched domains (DAC-2013, between Switchboard and NIST data both of which telephone data and English) (a) we propose an end-to-end DA method by adding the adversarial loss to the cross-entropy loss of the x-vector architecture, and (b) we evaluate our method on the challenging task of language adaptation. Moreover, we use Wasserstein GANs, a recently proposed version of GANs which addresses the vanishing gradient problem of GANs by replacing the domain discrepancy measure with Wasserstein distance \cite{gulrajani2017improved}. We also explore the uses of speaker labels in the adaptation data, even in the form of pseudo-labels for cases where the set is unlabeled, and we show that they are very helpful in attaining good performance. Finally, we examine the use of domain labels, which we concatenate to the layers of the network to learn domain-dependent transforms using a single network. To the best of our knowledge, we are the first to utilize domain labels in such a way.

\section{Domain adaptation in speaker recognition}
\label{sec:DASR}
During the past few years, several DA methods for speaker recognition have been proposed. In the case of unsupervised adaptation several methods apply a clustering algorithm in order to estimate speaker labels, with which a target-domain PLDA model is trained, while interpolation between the parameters of the source and target-domain PLDA models is applied to obtain the adapted PLDA \cite{garcia2014unsupervised,shum2014unsupervised,novoselov2014rbm}. The standard speaker recognition recipe in the Kaldi toolkit utilizes a simpler method for unsupervised adaptation that does not require clustering \cite{snyder_interspeech_2017}. This method aims at adjusting the covariance matrices of the PLDA model so that its total covariance better matches the total covariance of the adaptation data.  An alternative approach is to compensate for dataset shift in the i-vector space by modelling the subspace of dataset shift and removing those direction from the i-vectors \cite{aronowitz2014inter}. Other approaches do not attempt to cluster the utterances and perform DA simply by matching first and second order statistics of the i-vectors between source and target domains \cite{alam2018speaker}. 

Closer to the spirit of our work, two methods based on domain-adversarial adaptation and maximum mean discrepancy were recently introduced. In \cite{wang2018unsupervised}, a domain-adversarial neural networks (DANN) is employed in order to transform i-vectors to a domain-invariant representation space. The authors follow the recipe introduced in \cite{Ganin_JMLR2016} without using or estimating speaker labels for the target domain training data. They evaluate their method on DAC 2013 and demonstrate significant gains over other DA methods. In \cite{lin2018reducing}, DA is performed using maximum mean discrepancy (MMD) as a means to reduce the mismatch between the two distribution. The main differences compared to the DANN-based architecture in \cite{wang2018unsupervised} are (a) the use of MMD which makes training easier compared to GANs, (b) the use of reconstruction loss instead of cross entropy in the main network (i.e. an autoencoder architecture instead of a classifier over speakers \cite{makhzani2015adversarial}), and (c) the application field, which is the language adaptation task of NIST SRE16 instead of DAC 2013. The method yields slightly better results compared to inter-dataset variability compensation \cite{aronowitz2014inter}.   
\begin{figure}[!htbp]
\centering
\includegraphics[width=3.20in]{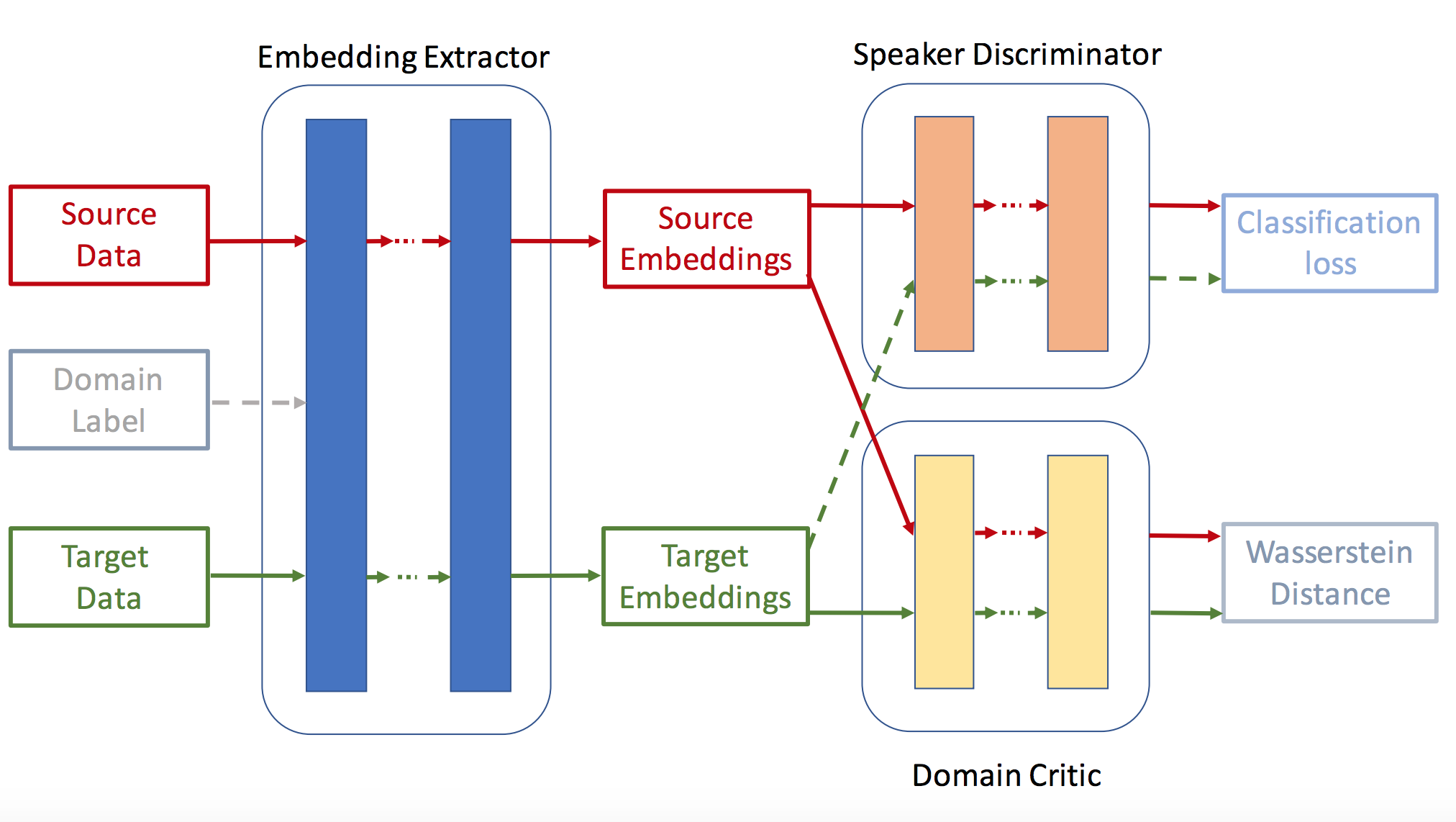}
\caption{Block-diagram of the architecture.}
\label{diagram}
\end{figure}

\section{Adversarial adaptation algorithm}
\label{sec:adv_adp_algo}

\subsection{Notation and Wasserstein distance}
We assume a labeled source dataset $X^s = \{(x_i^s,y_i^s) \}_{i=1}^{n_s}$ from the source domain $D_s$, and target dataset $X^t = \{(x_i^t,y_i^t) \}_{i=1}^{n_t}$ from the target domain $D_t$, where $x \in \mathbb{R}^{m,\cdot}$ denotes utterances and the labels $y_i^t$ may be given, estimated (e.g. using clustering) or not used at all. The two domains (i.e. languages) have different marginal data distributions, $P_{x^s}$ and $P_{x^t}$ respectively. The embedding extractor, which is a standard TDNN x-vector architecture up to the embedding layer implements a function $f_g : \mathbb{R}^{m,\cdot} \mapsto \mathbb{R}^d$ parametrized by $\theta_g$, where $d$ is the size of the embedding. The additional structure, useful only during training is called the domain critic and is a feed-forward neural network implementing a function $f_w : \mathbb{R}^{d} \mapsto \mathbb{R}$ parametrized by $\theta_w$. The Wasserstein distance between two representation distributions $P_{x^s}$ and $P_{x^t}$, where $h^s=f_g(x^s)$ and $h^t=f_g(x^t)$ is approximated by
\begin{equation}
    {\cal L}_{wd}(x^s,x^t) = \frac{1}{n^{s}}\sum_{x^s \in X^s}f_w(f_g(x^s)) - \frac{1}{n^{t}}\sum_{x^t \in X^t}f_w(f_g(x^t)).
\end{equation}
As proposed in \cite{gulrajani2017improved}, an improved method (compared to \cite{Arjovsky_icml2017}) for constraining $g_w$ to be 1-Lipschitz function (a necessary conditions so that ${\cal L}_{wd}$ is an approximation of the Wasserstein distance) is to introduce a gradient penalty loss
\begin{equation}
    {\cal L}_{grad}(\hat{h}) = \left(\norm{\nabla_{\hat{h}} f_w (\hat{h})}_2 - 1\right)^2,
\end{equation}
where $\hat{h}$ is a set of features created by randomly pairing and linearly combining features from $h^s$ and $h^t$ \cite{gulrajani2017improved}.

The speaker discriminator (i.e. the part of the x-vector network after the embedding layer) implements a function $f_c : \mathbb{R}^{d} \mapsto \mathbb{R}^l$, i.e. it maps the embeddings $h$ to the space of posterior probabilities over training speakers (either from source or from target domain) and it is parametrized by $\theta_c$ (separate linear and softmax layers are assumed for source and target domains). The classification loss ${\cal L}_c(x,y)$ is the standard cross-entropy over speakers. The architecture during training is illustrated in Fig. \ref{diagram}.
\begin{algorithm}
\caption{Domain adaptation algorithm}\label{DANN_alg}
\begin{algorithmic}[1]
\State Initialize feature extractor, domain critic, speaker discriminator $\theta_g, \theta_w, \theta_c$.
\State {\bf if} supervised $d \leftarrow (s,t)$ {\bf else} $d\leftarrow s$ {\bf end}
\State {\bf repeat}
\State $\,\,\,$ Sample minibatch $\{(x_i^s,y_i^s) \},\{(x_i^t,y_i^t) \}$
\State $\,\,\,$ {\bf for} $t = 1, \ldots, n$ {\bf do}
\State $\,\,\,$ $\,\,\,$ $h^s \leftarrow f_g(x^s)$, $h^t \leftarrow f_g(x^t)$
\State $\,\,\,$ $\,\,\,$ Sample $h$ as the random points along straight lines between $h^s$ and $h^t$ pairs.
\State $\,\,\,$ $\,\,\,$ $\hat{h}\leftarrow \{h^s,h^t,h\}$
\State $\,\,\,$ $\,\,\,$ $\theta_w \leftarrow \theta_w + \alpha_1 \nabla_{\theta_w}[{\cal L}_{wd}(x^s,x^t) - \gamma {\cal L}_{grad}(\hat{h})]$
\State $\,\,\,$ {\bf end for}
\State $\,\,\,$ $\theta_c \leftarrow \theta_c - \alpha_2 \nabla_{\theta_c} {\cal L}_c(x^d,y^d)$
\State $\,\,\,$ $\theta_g \leftarrow \theta_g - \alpha_2 \nabla_{\theta_g} [{\cal L}_c(x^d,y^d) + \delta {\cal L}_{wd}(x^s,x^t)] $
\State {\bf until} $\theta_g,\theta_w,\theta_c$ converge.
\end{algorithmic}
\end{algorithm}
\subsection{Training algorithm}
The training algorithm is given in Algorithm \ref{DANN_alg} (see \cite{Shen_aaai2018} for more information). As we observe, the critic $\theta_w$ tries to maximize ${\cal L}_{wd}$, i.e. to approximate the Wasserstein distance between the two domains while the feature extractor $\theta_g$ tries to minimize it, yielding the usual minimax optimization problem of GANs. In the inner loop of the algorithm, a set of points $\hat{h}$ are randomly chosen as linear combinations between randomly paired $h^s$ and $h^t$, on which the gradient penalization is applied, constraining $f_w$ to be a 1-Lipschitz function \cite{gulrajani2017improved}. Finally, when labels are used in the target domain, the classification loss is backpropagated for both sets.

\begin{table*}[!ht]
\caption{\label{tab:WGAN_Results} Results with adversarial adaptation. DCF refers to the average minDCF at the operating points 0.01 and 0.005. EER refers to equal error rate in \%. Cantonese and Tagalog are denoted by yue and tgl respectively.}
  \centerline{
    \begin{tabular}{l l r r r r r r r r } 
    \hline
     && \multicolumn{2}{c}{SRE18} &   \multicolumn{2}{c}{SRE16\_all} & \multicolumn{2}{c}{SRE16\_yue} & \multicolumn{2}{c}{SRE16\_tgl} \\
     && DCF  & EER & DCF & EER & DCF & EER & DCF  & EER\\
     \hline
     \hline
     \multirow{5}{*}{\rotatebox[origin=c]{90}{PLDA}}&
     Baseline   & 0.664 &10.01                 & 0.897   & 11.55& 0.553 & 7.28& 0.976 & 15.87\\ 
     \cline{2-10}
     &sup        & 0.652        &  9.59       & 0.859        & 10.94 & 0.536 & 6.79 & 0.950 & 15.19\\
     &adv        & 0.658         & 10.35       & 0.899        & 13.25 & 0.561 & 7.39    & 0.968 & 19.12\\
     &adv+sup    & 0.630        & 8.89        &  {\bf 0.737} & 9.88  & 0.501 & 5.73    & {\bf 0.855  }& 14.18\\
     &adv+lan+sup & {\bf 0.619} & {\bf 8.88} & 0.746        & {\bf 9.59} & {\bf 0.497} & {\bf 5.59} & 0.880 & {\bf 13.70}\\
    \hline 
   \end{tabular}
   }
  %}
\end{table*}

\subsection{Architecture and implementation details}
We use Tensorflow \cite{tensorflow2015-whitepaper} for implementing the adversarial adaptation.
We follow the standard Kaldi x-vector architecture \cite{snyder_interspeech_2017}, i.e. 5 TDNN layers with ReLU activation functions followed by batch normalization, then a pooling layer that accumulates mean and standard deviations, then two feed-forward layers with ReLU and batch normalization, then finally a softmax layer for classifying speakers. For the critic network we use two feed-forward layers with 512 units and leaky ReLU activation functions. The critic network takes the x-vector, i.e. the output of the first affine layer after pooling, as input and returns a scalar as output. The domain label is passed to the feature extractor as a binary variable which is concatenated to the input of every affine layer (0 corresponds to the source domain). This is equivalent to having domain-dependent biases, enabling the network to learn domain-dependent transforms.
Based on light tuning to make the training stable, we set the parameters of the adversarial training to $\gamma=10.0$, $\delta=0.1$, $\alpha_1=0.001$, $\alpha_2=1.0$, and $n= 10$. For supervised adaptation, we add a second softmax layer to the x-vector network, i.e. the source- and target-domain classifiers share all model parameters except those of their softmax layer. In order to better balance the source- and target-domain classification losses, we normalize them with the logarithm of their number of classes so that the loss of random prediction is approximately equal to one. After that, we set the weight for the target domain classification loss to 0.2 and the weight for the source domain classification loss to 0.8. 
In the experiments, we use minibatches containing 150 segments of the target domain data and 150 segments of the (labeled) source domain data. The lengths of the segments are 2-4s. We use stochastic gradient descent, starting with a learning rate of 1.0 which we then half every 5 epochs, where an epoch is defined to be 400 minibatches. We stop the training after 85 epochs. During the first 3 epochs, we trained only the critic and the network for source domain classification.

\section{Experiments}
\label{sec:expts}
We conduct experiments on two databases, the NIST SRE 2018 {\it cmn2} evaluation set using the unlabeled {\it cmn2} development set for adaptation, and the NIST SRE16 evaluation set \cite{NIST_SRE16} using the unlabeled major data set for adaptation. It should be noted that the SRE16 evaluation data as well as the unlabeled major data contains utterances from two languages (Cantonese and Tagalog) which is not ideal for our proposed methods, since language labels are not provided (and are not estimated by our algorithm). Hence, we treat two distinct languages as one, which is clearly suboptimal. Further, for the SRE18 unlabeled development set, we have access to the telephone numbers which should help the supervised training compared to just using utterance IDs. The adaptation data is augmented similarly to the training data \cite{Kaldi_xvector16}, i.e. with babble, noise, music and reverberated versions of the utterances.

The baseline x-vector model is trained by the Kaldi toolkit \cite{povey2011kaldi} using the Kaldi SRE16 x-vector recipe \cite{Kaldi_xvector16} but with additional training data from voxceleb2, resulting in 12170 training speakers. We apply the adversarial DA on this model rather than training a model with adversarial DA from the beginning. %(Training a baseline model with Tensorflow generally seems to work a well as with Kaldi but since we already had an available Kaldi model trained on the particular training data we used it). 
We experimented with two backends. The first is an identical backend to the one in the Kaldi x-vector recipe \cite{Kaldi_xvector16}. This backend involves a preprocessing step which first reduces the x-vector dimension by LDA from 512 to 150, and then applies an unusual variant of length-norm\footnote{See https://github.com/kaldi-asr/kaldi/blob/master/src/ivector/plda.cc}. %This backend was implemented in python based on our in-house toolkit {\it Pytel}. We refer to this backend as PLDA.
%The second backend is the heavy tailed (HT)-PLDA backend proposed in \cite{SilnovaIS18}. This backend has the advantage that no preprocessing of the x-vectors is needed for good performance. %This backend was implemented in Matlab. We refer to this backend as HT-PLDA.
In all experiments, we center the evaluation data at the mean of the unlabeled development set instead of the mean of the PLDA training set. This can be seen as a light form of adaptation. 

\subsection{Results with adversarial adaptation}
\label{sec:exp_adv_adp}
In this subsection, we evaluate the different variants of adversarial DA detailed in Section \ref{sec:adv_adp_algo} with Gaussian PLDA backend. In summary, three observations can be made. First, adversarial adaptation is effective when it is combined with supervised training on the target data. Without supervised training on the target data adversarial adaptation degrades the performance. 
Second, including language labels as side-information to the TDNN layers helps for SRE18 but not SRE16. This is not too surprising, considering that SRE16 contains two languages which we treat as one. 
\subsection{Interaction with backend adaptation}
%In this subsection, we compare adversarial DA with backend DA as well as with the combination of adversarial and backend DA. 
We present result for the standard Kaldi-style unsupervised PLDA DA.
This method essentially estimates the excess variance in the adaptation data and distributes a portion, $\xi$, of it to the PLDA between-class covariance matrix and a portion, $\eta$, to the PLDA within-class covariance matrix\footnote{See https://github.com/kaldi-asr/kaldi/blob/master/src/ivector/plda.cc}. (Our experimentation with supervised PLDA adaptation using clustering methods were unsuccessful for both for the baseline and the adversarial DA model.) 
In Kaldi $\xi=1-\eta$ and in the SRE16 x-vector recipe \cite{Kaldi_xvector16}, $\xi=0.25$. The results for these settings are shown in the first part of Table \ref{tab:WGAN_Results_adp} (PLDA adp 1). As can be seen, adversarial DA degrades the performance. However, the Kaldi settings of $\xi$ and $\eta$ may not be optimal when adversarial DA is applied. For example, if the adversarial DA manages to remove between-class variability much better than within-class variability, the balance between $\xi$ and $\eta$ should be different. Moreover, in these experiments we use the same adaptation data for both the adversarial DA and the PLDA adaptation. After being used for adversarial DA, the adaptation data is most likely closer to the source data than what unseen (test) data is, meaning that the PLDA adaptation will not be strong enough. To mitigate this, we tune $\xi$ and $\eta$ on the SRE18 labeled development set (without requiring that they sum to 1). The results are shown in the second part of Table \ref{tab:WGAN_Results_adp} (PLDA adp 2). As we observe, tuning $\xi$ and $\eta$ helps both the baseline and the models from adversarial training for SRE18. In terms of EER, the adversarial training now complements PLDA DA. For SRE16 using these values of $\xi$ and $\eta$ is worse than using the original Kaldi settings. Probably, the SRE18 development set is not similar enough to SRE16 for $\xi$ and $\eta$ to be properly estimated.

\begin{table}[!ht]
\caption{\label{tab:WGAN_Results_adp} Results with adversarial DA and PLDA DA.}
  %\centerline{
    \begin{tabular}{l l r r r r } 
    \hline
     && \multicolumn{2}{c}{SRE18} &   \multicolumn{2}{c}{SRE16\_all} \\ 
     && DCF  & EER & DCF  & EER \\
    \hline
    \hline 
    \multirow{5}{*}{\rotatebox[origin=c]{90}{PLDA adp 1}}&
    Baseline      &  0.580&	9.05& 0.613& 8.01  \\ 
     \cline{2-6}
     &sup         & { \bf 0.576 }&	{ \bf 8.90}& { \bf 0.611}& { \bf 7.74}\\
     &adv          & 0.616&	9.70& 0.664& 9.42\\          
     &adv+sup     & 0.591&	9.15& 0.630&	8.10\\  
     &adv+lan+sup & 0.588&	9.03& 0.615&	7.92\\ 
    \hline
    \hline
    \multirow{5}{*}{\rotatebox[origin=c]{90}{PLDA adp 2}}&
     Baseline      & 0.572 & 8.51 &  0.656&	7.98   \\ 
     \cline{2-6}
     &sup         &{ \bf 0.567} & 8.67 & {\bf 0.624 }&	{\bf 7.66}        \\
     &adv          & 0.602& 9.21 & 0.677&	9.15         \\          
     &adv+sup     &0.578 & 8.28 & 0.649& 8.00  \\  
     &adv+lan+sup & 0.576& {\bf 8.25} & 0.651&8.00 \\ 
     \hline
    \hline 
%    \multirow{5}{*}{\rotatebox[origin=c]{90}{HT-PLDA}}&
%      Baseline      & 0.621	&8.664  & 0.942 &12.96    \\ 
%     \cline{2-6}
%     &sup.         & 0.614	&8.532  & 0.907	&12.20         \\
%     &adv          & 0.618	&9.098  & 0.875	&12.90    \\          
%     &adv+sup.     & 0.593	&8.456  & 0.823	&11.16 \\  
%     &adv+lan+sup. & 0.588	&8.331  & 0.815	&11.22 \\ 
%     \hline
%    \hline 
   \end{tabular}
   %}
  %}
  \vspace{-0.4cm}
\end{table}

%For HT-PLDA we experimented with speaker clustering followed by training a target domain PLDA model and interpolation of the model parameters in the case of SRE16, and using the telephone numbers as speaker IDs in the case of SRE18. With this backend it is beneficial to add adversarial DA. Our best result on SRE18 is obtained with this adaptation on top of x-vectors obtained from adv+lan+sup adaptation (EER: 6.90\%, DC1: 0.498, DC2: 0.558). 
\vspace{-0.3cm}
\subsection{Adapting only the first layer after pooling}
In the main experiment we adapted all layers of the network in an end-to-end fashion. In Table \ref{tab:WGAN_Results_no_e2e}, we show results of adapting only the first layer after pooling. This is similar to i-vector AD approach in \cite{Wang_icassp2018}, although, here the transformation we learn is an affine dimensionality reduction of x-vectors.
\begin{table}[!ht]
\caption{\label{tab:WGAN_Results_no_e2e} Results with adversarial DA after pooling.}
  %\centerline{
    \begin{tabular}{l l r r r r } 
    \hline
     && \multicolumn{2}{c}{SRE18} &   \multicolumn{2}{c}{SRE16\_all} \\ 
     && DCF  & EER & DCF  & EER \\
    \hline
    \hline
    \multirow{5}{*}{\rotatebox[origin=c]{90}{PLDA}}&
     Baseline      &0.664 & 10.01 & 0.897 &11.55    \\ 
     \cline{2-6}
     &sup         & 0.667& 9.92 &0.887 &11.53         \\
     &adv          & {\bf 0.615}& 9.03& 0.751 &9.91         \\          
     &adv+sup     & 0.629& {\bf 8.92} & 0.741 &9.79 \\  
     &adv+lan+sup & 0.620& 9.01 & {\bf 0.726} & {\bf 9.63} \\ 
     \hline
    \hline 
%    \multirow{5}{*}{\rotatebox[origin=c]{90}{HT-PLDA}}&
%      Baseline     &0.668 & 9.74 & 0.847 & 12.30 \\ 
%     \cline{2-6}
%     &sup.         & 0.659& 9.64 & 0.836 &12.12 \\
%     &adv          & 0.642& 9.02 & 0.827 &11.28 \\          
%     &adv+sup.     & 0.643& 8.98 & 0.818 &10.61 \\  
%     &adv+lan+sup. & 0.639& 9.05 & 0.801 & 10.40 \\ 
%     \hline
%    \hline 
   \end{tabular}
   %}
  %}
  \vspace{-0.4cm}
\end{table}

Two observations can be made. First, when adapting only this layer there is no clear advantage of neither supervised adaptation nor including language labels. Second, in SRE16 adapting only this layer performs similar to adapting all layers. %We did some experiments were we adapted the first layer after pooling as well as some of the layers befor pooling. 
This can possibly be mitigated by some form of regularization in the earlier layers. %, but we leave it for future work.

\section{Conclusion and future work}
In this paper we introduced an end-to-end DA method for x-vectors based on Wasserstein GANs. We examined several configurations, especially with respect to the use of speaker and domain labels. We provided a detailed evaluation on NIST SRE16 and SRE18, and a fair comparison with state-of-the-art DA methods. The results show the effectiveness of the method in certain experiments, but also emphasize the need for further experimentation. To this end, we consider training the system from scratch with adversarial loss, applying the method to other DA tasks such as gender and channels, as well as addressing overfitting to the target domain data.

\vfill\pagebreak

% References should be produced using the bibtex program from suitable
% BiBTeX files (here: strings, refs, manuals). The IEEEbib.bst bibliography
% style file from IEEE produces unsorted bibliography list.
% -------------------------------------------------------------------------
\bibliographystyle{IEEEbib}
\bibliography{strings,refs}

\end{document}